\documentclass[11pt]{article}

\PassOptionsToPackage{hyperfootnotes=false}{hyperref}
\PassOptionsToPackage{table}{xcolor}
\usepackage[preprint]{acl}

\usepackage{times}
\usepackage{latexsym}

\usepackage[T1]{fontenc}

\usepackage[utf8]{inputenc}

\usepackage{microtype}

\usepackage{inconsolata}

\usepackage{graphicx}
\usepackage{booktabs}
\usepackage{multirow}
\usepackage{enumitem}
\usepackage{etoolbox}
\apptocmd{\thebibliography}{\setlength{\itemsep}{0pt}\setlength{\parsep}{0pt}}{}{}
\usepackage{amsmath}
\usepackage[nointegrals]{wasysym}
\usepackage{tabularx}
\usepackage{longtable}
\usepackage{array}
\usepackage{pifont}
\usepackage{amssymb}
\usepackage{adjustbox}
\newcolumntype{L}{>{\raggedright\arraybackslash}X}

\usepackage{tikz}
\usetikzlibrary{arrows.meta}
\usepackage[edges]{forest}
\usepackage{makecell}
\usepackage{mydef}
\title{A Survey on Long-Term Memory Security in LLM Agents:\\ Attacks, Defenses, and Governance Across the Memory Lifecycle}

\author{
\textbf{Zehao Lin}$^{1}$\thanks{Equal Contribution.}, 
\textbf{Xixuan Hao}$^{1,2}$$^*$,
\textbf{Renyu Fu}$^{1,4}$, 
\textbf{Shaobo Cui}$^{3}$\\
\textbf{Kai Chen}$^{1}$\textbf{,}
\textbf{Chunyu Li}$^{1}$\textbf{,}
\textbf{Zhiyu Li}$^{1}$\thanks{Corresponding Author.}\textbf{,}
\textbf{Feiyu Xiong}$^{1}$\\
$^1$  MemTensor (Shanghai) Technology Co., Ltd.\\
$^2$ The Hong Kong University of Science and Technology (Guangzhou)\\
$^3$ Shanghai Jiao Tong University\quad $^4$ Peking University\\
\texttt{\{linzh, fury, chenk, licy, lizy, xiongfy\}@memtensor.cn},\\
\texttt{xhao390@connect.hkust-gz.edu.cn}, 
\texttt{shaobo.cui@sjtu.edu.cn},\\
\texttt{furenyu26@stu.pku.edu.cn}
}

\newcommand{\grayline}{%
  \arrayrulecolor{black!20}\hline
  \arrayrulecolor{black}%
}

\hypersetup{
  pdftitle={A Survey on Long-Term Memory Security in LLM Agents: Attacks, Defenses, and Governance Across the Memory Lifecycle},
  pdfauthor={Zehao Lin; Xixuan Hao; Renyu Fu; Shaobo Cui; Kai Chen; Chunyu Li; Zhiyu Li; Feiyu Xiong},
  pdfsubject={Long-term memory security in LLM agents: attacks, defenses, and governance},
  pdfkeywords={LLM agents, long-term memory, memory security, memory poisoning, provenance, rollback, verified forgetting},
  pdfdisplaydoctitle=true
}

\begin{document}
\raggedbottom
\maketitle
\begin{abstract}
The emergence of writable, cross-session persistent memory in LLM agents introduces a qualitatively different threat landscape from conventional input-centric security concerns, characterized by three properties: persistence, statefulness, and propagation. 
To systematically characterize this landscape, we propose a Memory Lifecycle Framework that organizes attacks, defenses, and their cross-phase dependencies along two axes: six lifecycle phases (\textsc{Write}, \textsc{Store}, \textsc{Retrieve}, \textsc{Execute}, \textsc{Share \& Propagate}, \textsc{Forget \& Rollback}) and four security objectives (Integrity, Confidentiality, Availability, Governance).
This analysis in turn exposes the need for explicit security requirements at the system level, motivating \textit{Verifiable Memory Governance} (VMG), a framework of five architectural primitives for specifying and evaluating auditable, recoverable control over memory state within declared system and threat-model boundaries.
Our analysis motivates integrating provenance, versioning, and policy-aware retention with retrieval and execution defenses to support auditable prevention, containment, and recovery across the lifecycle.

\end{abstract}

\noindent{\small\textit{Expanded arXiv version, 19 November 2026. Revision notes appear in Appendix~\ref{app:revision}.}}

\section{Introduction}\label{sec:intro}

Large language models (LLMs) are increasingly serving as the foundation
for interactive and autonomous agents~\cite{zhang2024memorysurvey,wang2024survey}.
As their role expands from generating responses to supporting decisions,
tool use, and user interactions, their security risks also extend
beyond isolated outputs~\cite{zhan2024injecagent,zhang2025asb,greshake2023not}.
In recent years, research on the security of LLM-based agents has traditionally centered on two
areas: knowledge encoded parametrically in model weights~\cite{carlini2021extracting,bourtoule2021machine,maini2024tofu} and prompt manipulation within single- or multi-turn interactions~\cite{liu2024formalizing,russinovich2025crescendo}.
As these agents evolve from stateless chatbots into autonomous systems
equipped with persistent long-term memory, however, the focus of security
concern is shifting from transient inputs to durable state.
This transition qualitatively changes the landscape of security:
attackers no longer merely seek to influence an immediate response, but
to manipulate how the system \emph{remembers, retrieves, plans, and acts}
across future interactions.

Memory matters because distinctions preserved from past experience can improve later decisions. \citet{lin2026value} distinguish the decision value available in a history, retained in persistent state, and realized by a bounded reader under specified demands and losses, while accounting for recovery and revision costs. For this survey, that perspective motivates protecting both authorized use and useful continuity: attacks can corrupt retained support or its use, while defenses can discard legitimate support. The design problem therefore connects utility, cost, and governance: what merits retaining, who may access or revise it, and how interventions affect future work.

Existing prompt-injection and RAG-corruption mitigations are fundamentally limited for Long-Term Memory (LTM) security: 
they largely operate within a single session or retrieval episode~\cite{debenedetti2025camel,xiang2024robustrag}
and do not fully address 
the cross-session, stateful nature of persistent memory~\cite{packer2023memgpt,zhong2024memorybank,chhikara2025mem0,zou2026etamp}.
Increasing the context-window size, for instance, does not subsume the security
challenges introduced by LTM.
Although a larger context window improves within-session recall, it does not create persistent memory, as its contents are not automatically retained, shared, or retrieved across future interactions~\cite{packer2023memgpt,zhong2024memorybank,chhikara2025mem0}. 
Provenance tagging, principal-scoped retrieval, rollbackable state,
and verified forgetting arise whenever content outlives a session or crosses a principal boundary~\cite{provdm2013,li2025memos-full,rezazadeh2025collaborative,bourtoule2021machine}.

Through a structured review of existing literature, we identify three properties of long-term memory that require analysis beyond isolated inputs and outputs.
(1) \textbf{Persistence}. A poisoned entry can be recalled across an
indefinite number of future sessions, long after the originating
context has closed~\citep{dong2025minja,zou2026etamp}.
(2) \textbf{Statefulness}. The unit of security analysis shifts from isolated input instances to the agent's evolving memory state.
An agent that has accumulated a cluster of subtly biased episodic
memories may exhibit behavioral drift long before any single memory
entry triggers a conventional safety classifier~\citep{dong2025minja,zou2026etamp}.
(3) \textbf{Propagation}. In multi-agent and shared-state systems,
contamination spreads through inter-agent messages, shared stores, and
tool arguments, producing cascading effects across session, role, and
user boundaries~\citep{gu2024agentsmith,men2024contagious,elyagoubi2026agentleak}.

Taken together, these properties suggest that LTM is not merely an extension of context or storage capacity, but a new security substrate that changes what must be protected, monitored, and governed. 
Agent-security frameworks organized around individual interfaces provide only a partial account of these risks; broader architectural and data-centric surveys offer complementary views of cross-session behavior~\citep{chu2026layered,lahjouji2026knowtoomuch}.
The central question we pursue is:
\emph{when an LLM agent acquires writable, retrievable, cross-session
persistent memory, what qualitative change occurs in the security
landscape, and how should the field organize its response?}

\begin{table}[t]
\centering
\caption{LTM security coverage in related surveys. Lifecycle: explicit security analysis across memory operations; Benign Risk: unintended disclosure or drift; Framework: dedicated memory-governance requirements.
{\color[HTML]{2191A8}\ding{51}}~=~explicit;
{\color[HTML]{2191A8}\LEFTcircle}~=~partial;
{\color[HTML]{E7524C}\ding{55}}~=~not addressed.}
\label{tab:recent_survey_new}
\begin{adjustbox}{max width=\columnwidth}
\footnotesize
\setlength{\tabcolsep}{4pt}
\begin{tabular}{@{}lccc@{}}
\toprule
\textbf{Survey} &
\textbf{Lifecycle} &
\textbf{Benign Risk} &
\makecell{\textbf{LTM Security}\\\textbf{Framework}} \\
\midrule
\citet{zhang2024memorysurvey}      & {\color[HTML]{E7524C}\ding{55}}  & {\color[HTML]{E7524C}\ding{55}} & {\color[HTML]{E7524C}\ding{55}} \\
\citet{wu2025humanai}              & {\color[HTML]{E7524C}\ding{55}}  & {\color[HTML]{E7524C}\ding{55}} & {\color[HTML]{E7524C}\ding{55}} \\
\citet{liang2025aimeetsbrain}      & {\color[HTML]{E7524C}\ding{55}}  & {\color[HTML]{E7524C}\ding{55}} & {\color[HTML]{E7524C}\ding{55}} \\
\citet{he2025emergedsecurity}      & {\color[HTML]{E7524C}\ding{55}}  & {\color[HTML]{E7524C}\ding{55}} & {\color[HTML]{E7524C}\ding{55}} \\
\citet{mu2026securerag}            & {\color[HTML]{2191A8}\LEFTcircle} & {\color[HTML]{E7524C}\ding{55}} & {\color[HTML]{E7524C}\ding{55}} \\
\citet{bodea2026sokragprivacy}     & {\color[HTML]{2191A8}\LEFTcircle} & {\color[HTML]{E7524C}\ding{55}} & {\color[HTML]{E7524C}\ding{55}} \\
\citet{hu2025memoryage}            & {\color[HTML]{E7524C}\ding{55}}  & {\color[HTML]{E7524C}\ding{55}} & {\color[HTML]{E7524C}\ding{55}} \\
\citet{tang2026securityllmagents}  & {\color[HTML]{2191A8}\LEFTcircle} & {\color[HTML]{E7524C}\ding{55}} & {\color[HTML]{E7524C}\ding{55}} \\
\citet{luo2026storage}             & {\color[HTML]{2191A8}\LEFTcircle} & {\color[HTML]{E7524C}\ding{55}} & {\color[HTML]{E7524C}\ding{55}} \\
\citet{chu2026layered}            & {\color[HTML]{2191A8}\ding{51}} & {\color[HTML]{2191A8}\ding{51}} & {\color[HTML]{2191A8}\LEFTcircle} \\
\citet{lahjouji2026knowtoomuch}    & {\color[HTML]{2191A8}\LEFTcircle} & {\color[HTML]{2191A8}\ding{51}} & {\color[HTML]{2191A8}\LEFTcircle} \\
\citet{chen2026saver}            & {\color[HTML]{2191A8}\ding{51}} & {\color[HTML]{2191A8}\ding{51}} & {\color[HTML]{2191A8}\LEFTcircle} \\
\midrule
\textbf{Ours} &
{\color[HTML]{2191A8}\ding{51}} &
{\color[HTML]{2191A8}\ding{51}} &
{\color[HTML]{2191A8}\ding{51}} \\
\bottomrule
\end{tabular}
\end{adjustbox}
\end{table}

\noindent
\textbf{Comparison with existing surveys.}
Table~\ref{tab:recent_survey_new} compares the cited surveys' explicit treatment of lifecycle security, benign memory risks, and governance frameworks. Our work connects these dimensions within one analysis.
\citet{chu2026layered} organizes agent security by architectural layer and threat timescale, including memory consolidation and drift. \citet{lahjouji2026knowtoomuch} organize privacy by data surface and analyze cross-session inference, consent, and revocation. SAVER follows reusable state through adaptation, exposure, and repair across memory, skills, workflows, and model updates~\citep{chen2026saver}. Focusing on individual memory operations lets us specify who may retrieve a derived record, what a rollback must restore, and which channels to probe after deletion.

\noindent
\textbf{Our contributions.}
Through a structured literature review with targeted updates, we provide a taxonomy framework for LTM security in LLM agents, as shown in Figure~\ref{fig:taxonomy}.
This taxonomy serves as an organizational scaffold for analyzing how memory-related attacks, defenses, and governance requirements are connected across the memory lifecycle.
In summary, our work makes three contributions:
\begin{itemize}[leftmargin=*]
    \item \textit{A Memory Lifecycle Framework as an Organizational Principle}.
We propose a six-lifecycle-phase framework
(\textsc{Write}; \textsc{Store}; \textsc{Retrieve}; \textsc{Execute}; \textsc{Share \& Propagate}; \textsc{Forget \& Rollback}), exposing cross-phase attack chains invisible to single-turn or input-centric frameworks.
\item
\textit{A Structured Survey of Attacks and Defenses for LTM}. We map representative attacks and defenses onto the lifecycle framework, identifying uneven coverage across phases and gaps in evaluating how prevention, containment, and recovery work together across sessions and principal boundaries.

\item
\textit{A Requirements Framework for Verifiable Memory Governance (VMG)}.
We formalize memory governance into five primitives (Write Authorization,
Provenance Visibility, Principal-Scoped Retrieval, Rollbackability,
Verified Forgetting), each with a predicate definition and evaluation
metric. These criteria make governance requirements explicit and guide
audits within stated system and threat-model boundaries.
\end{itemize}

\definecolor{mycolor}{RGB}{215, 245, 200}
\definecolor{middlecolor}{HTML}{EBF1FA}

\tikzstyle{leaf}=[
    draw=hiddendraw,
    rounded corners,
    minimum height=1em,
    fill=mycolor,
    text opacity=1, 
    align=center,
    fill opacity=.5,  
    text=black,
    align=left,
    font=\scriptsize,
    inner xsep=3pt,
    inner ysep=1pt,
]
\tikzstyle{middle}=[
    draw=hiddendraw,
    rounded corners,
    minimum height=1em,
    fill=middlecolor,
    text opacity=1, 
    align=center,
    fill opacity=1,  
    text=black,
    align=left,
    font=\scriptsize,
    inner xsep=3pt,
    inner ysep=1pt,
]

\begin{figure*}[p]
    \vspace{-1em}
    \centering
    \begin{adjustbox}{max width=\textwidth,max totalheight=0.90\textheight}
    \begin{forest}
        for tree={
            forked edges,
            grow=east,
            reversed=true,
            anchor=base west,
            parent anchor=east,
            child anchor=west,
            base=middle,
            font=\footnotesize,
            rectangle,
            line width=0.7pt,
            draw=output-black,
            rounded corners,
            align=left,
            minimum width=2em,
            s sep=3pt,
            inner xsep=3pt,
            inner ysep=1pt,
        },
        where level=1{text width=5.5em}{},
        where level=2{text width=6.5em,font=\footnotesize}{},
        where level=3{font=\footnotesize}{},
        where level=4{font=\footnotesize}{},
        where level=5{font=\footnotesize}{},
        [LTM Security\\in LLM Agents, middle, rotate=90, font=\small, anchor=north, edge=output-black
            [Threats and\\Attack Surfaces, middle, edge=output-black, text width=5.3em
                [Write, middle, text width=7em, edge=output-black
                    [Corpus-level Poisoning: AgentPoison~\citep{chen2024agentpoison}{,} \\BadRAG~\citep{xue2024badrag}{,} Phantom~\citep{chaudhari2024phantom}, leaf, text width=22em, edge=output-black]
                    [Query-induced Injection: MINJA~\citep{dong2025minja}{,} \\InjecMEM~\citep{tian2026injecmem}, leaf, text width=22em, edge=output-black]
                    [Environment-injected: eTAMP~\citep{zou2026etamp}{,} \\SpAIware~\citep{herrador2025spaiware}, leaf, text width=22em, edge=output-black]
                    [Procedural Grafting: MemoryGraft~\citep{srivastava2025memorygraft}, leaf, text width=22em, edge=output-black]
                ]
                [Store, middle, text width=7em, edge=output-black
                    [Compression Interfaces: A-MEM~\citep{xu2025amem}{,} \\MemGPT~\citep{packer2023memgpt}, leaf, text width=22em, edge=output-black]
                    [Provenance Interfaces: MemOS~\citep{li2025memos-full}, leaf, text width=22em, edge=output-black]
                ]
                [Retrieve \& Execute, middle, text width=7em, edge=output-black
                    [Control-flow Hijacking: MCFA~\citep{xu2026mcfa}, leaf, text width=22em, edge=output-black]
                    [Retrieval Poisoning: PoisonedRAG~\citep{zou2025poisonedrag}{,} \citep{liang2025graphrag}{,} \\KEPo~\citep{chen2026kepo}, leaf, text width=22em, edge=output-black]
                    [Backdoor / Trigger: AgentPoison~\citep{chen2024agentpoison}{,} \\Phantom~\citep{chaudhari2024phantom}, leaf, text width=22em, edge=output-black]
                ]
                [Share \& Propagate, middle, text width=7em, edge=output-black
                    [Cross-agent Contagion: Agent Smith~\citep{gu2024agentsmith}{,} \citep{men2024contagious}, leaf, text width=22em, edge=output-black]
                    [Self-replicating Worms: AI Worm~\citep{cohen2024morrisii}, leaf, text width=22em, edge=output-black]
                    [Cross-user Contamination: \citep{yang2026ucc}{,} \citep{wang2025mextra}, leaf, text width=22em, edge=output-black]
                ]
                                [Forget \& Rollback, middle, text width=7em, edge=output-black
                    [Residual Derivatives: Ghost of the Past~\citep{zhang2024ghost}{,} \\PersistBench~\citep{pulipaka2026persistbench}, leaf, text width=22em, edge=output-black]
                    [Reappearance after Deletion: \citep{wang2024ragunlearning}{,} \\Agentic Unlearning~\citep{wang2026sbu}, leaf, text width=22em, edge=output-black]
                    [Recovery / Traceback Interfaces: RAGForensics~\citep{zhang2025ragforensics}{,} \\ MemOS~\citep{li2025memos-full}, leaf, text width=22em, edge=output-black]
                ]
            ]
            [Defenses, middle, edge=output-black, text width=5.3em
                [Write-time Prevention, middle, text width=7em, edge=output-black
                    [Human-verified Freeze: VerificAgent~\citep{nguyen2025verificagent}, leaf, text width=22em, edge=output-black]
                    [Provenance Tagging: PROV-AGENT~\citep{souza2025provagent}, leaf, text width=22em, edge=output-black]
                ]
                [Retrieve-time Detection, middle, text width=7em, edge=output-black
                    [Certifiable Aggregation: RobustRAG~\citep{xiang2024robustrag}, leaf, text width=22em, edge=output-black]
                    [Trust Scoring: TrustRAG~\citep{zhou2025trustrag}{,} SeCon-RAG~\citep{si2025seconrag}, leaf, text width=22em, edge=output-black]
                    [Activation-based: RevPRAG~\citep{tan2025revprag}, leaf, text width=22em, edge=output-black]
                    [Memory-native Consensus: A-MemGuard~\citep{wei2025amemguard}, leaf, text width=22em, edge=output-black]
                ]
                [Execute-time Containment, middle, text width=7em, edge=output-black
                    [Info-flow Control: CaMeL~\citep{debenedetti2025camel}{,} \\FIDES~\citep{costa2025fides}{,} PCAS~\citep{palumbo2026pcas}, leaf, text width=22em, edge=output-black]
                    [Tool Privilege: Progent~\citep{shi2025progent}{,} \\IsolateGPT~\citep{wu2024isolategpt}{,} \citep{jacob2025typeprivilege}, leaf, text width=22em, edge=output-black]
                    [Contextual Privacy: \citep{wen2026cdi}{,} AgentSpec~\citep{wang2025agentspec}, leaf, text width=22em, edge=output-black]
                ]
                [Share-time Governance, middle, text width=7em, edge=output-black
                    [Principal-scoped Access: \citep{rezazadeh2025collaborative}{,} \citep{li2025aac}, leaf, text width=22em, edge=output-black]
                    [Contagion Detection: BlindGuard~\citep{miao2025blindguard}, leaf, text width=22em, edge=output-black]
                ]
                [Forget/Rollback Recovery, middle, text width=7em, edge=output-black
                    [Forensic Traceback: RAGForensics~\citep{zhang2025ragforensics}, leaf, text width=22em, edge=output-black]
                    [Snapshot / Rollback: MemOS~\citep{li2025memos-full}, leaf, text width=22em, edge=output-black]
                    [Machine Unlearning: \citep{bourtoule2021machine}{,} TOFU~\citep{maini2024tofu}{,} \\RAG-unlearn~\citep{wang2024ragunlearning}, leaf, text width=22em, edge=output-black]
                ]
            ]
                        [Verifiable Memory\\Governance, middle, edge=output-black, text width=5.3em
                [Write Authorization: VerificAgent~\citep{nguyen2025verificagent}{,} MINJA~\citep{dong2025minja}{,} eTAMP~\citep{zou2026etamp}, leaf, text width=30.8em, edge=output-black]
                [Provenance Visibility: MemOS~\citep{li2025memos-full}{,} PROV-AGENT~\citep{souza2025provagent}, leaf, text width=30.8em, edge=output-black]
                [Principal-scoped Retrieval: \citep{rezazadeh2025collaborative}{,} \citep{li2025aac}{,} \\ NIST AI RMF~\citep{nist2023airmf}, leaf, text width=30.8em, edge=output-black]
                [Rollbackability: ChronoMem~\citep{su2026chronomem}{,} RAGForensics~\citep{zhang2025ragforensics}, leaf, text width=30.8em, edge=output-black]
                [Verified Forgetting: TOFU~\citep{maini2024tofu}{,} RAG-unlearn~\citep{wang2024ragunlearning}, leaf, text width=30.8em, edge=output-black]
            ]
        ]
    \end{forest}
    \end{adjustbox}
    \caption{Representative taxonomy of long-term memory security. System examples identify architectural attack surfaces, rather than demonstrated attacks on each system. Table~\ref{tab:new-evidence} relates lifecycle claims to their observables and evaluation conditions.
    }
    \vspace{-1em}
    \label{fig:taxonomy}
\end{figure*}

\section{Memory Lifecycle Framework}\label{sec:framework}

\label{ssec:phases}

Attacks and defenses are often studied at particular interfaces, while surveys organize them by memory architecture, threat type, or agent layer~\cite{zhang2024memorysurvey,wu2025humanai,he2025emergedsecurity,tang2026securityllmagents,mu2026securerag,chu2026layered}. A lifecycle view connects these perspectives: poisoning implanted during the Write stage may be activated during the Retrieve stage and exploited during the Execute stage, while the most effective point of defensive intervention may lie in a different phase.
We organize our analysis around these transitions to identify where attacks are seeded, how they propagate, and where defenses can intervene.

\begin{figure*}[t]
\centering
\includegraphics[width=\textwidth]{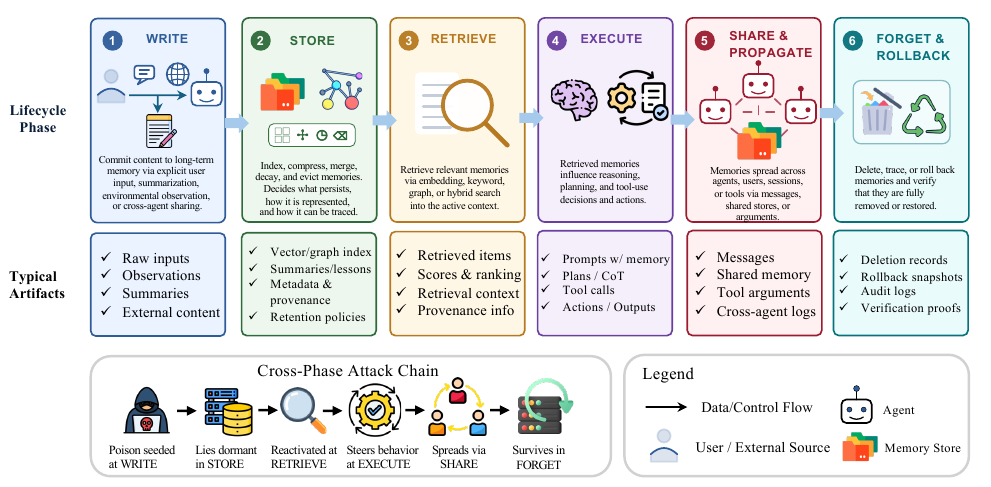}
\vspace{-2em}
\caption{The six-phase memory lifecycle of an LLM agent.
Each phase introduces distinct security questions.
}
\vspace{-1em}
\label{fig:lifecycle}
\end{figure*}

As shown in Figure~\ref{fig:lifecycle}, we decompose the memory lifecycle into six phases: \textit{Write, Store, Retrieve, Execute, Share \& Propagate, Forget \& Rollback}.
\ding{182} \textbf{Write}. It commits content to long-term memory through explicit
user instruction, implicit dialogue summarization, environmental observation,
or cross-agent memory sharing. 
The central security question is whether memory writes are subject to explicit
source authentication and authorization, rather than allowing untrusted
external content to enter memory as if it were user-endorsed~\cite{dong2025minja,zou2026etamp,tian2026injecmem,greshake2023not}.
\ding{183} \textbf{Store}. This phase indexes, compresses,
merges, decays, and evicts written content~\cite{packer2023memgpt,zhong2024memorybank,xu2025amem,chhikara2025mem0,li2025memos-full}. It determines whether poisoned
memory can persist, remain traceable, or become amplified. When a poisoned
entry is distilled into a high-level lesson, it may be effectively promoted
to higher retrieval priority and presented with greater apparent authority.
\ding{184} \textbf{Retrieve}. This phase brings previously stored memory
entries or encoded states back into use. Text and graph systems select evidence through embedding similarity, keyword matching,
graph-based retrieval, or hybrid routing~\cite{lewis2020rag,edge2024graphrag,zhang2024memorysurvey}.
Rather than serving as a neutral
lookup mechanism, retrieval shapes downstream reasoning by selecting which
memories enter the model context and which are excluded~\cite{barnett2024seven,chen2024agentpoison,zou2025poisonedrag,srivastava2025memorygraft}.
\ding{185} \textbf{Execute}. In this phase, retrieved memories begin to shape
the model's planning, reasoning, and tool-use decisions~\cite{sumers2024coala,packer2023memgpt,wang2024survey}. Recent work on memory
control-flow attacks~\citep{xu2026mcfa} shows that a sufficiently salient
retrieved memory can override explicit user instructions, thereby shifting
memory poisoning from a data-integrity issue to a control-flow problem.
\ding{186} \textbf{Share \& Propagate}. This phase determines whether
contamination spreads laterally, from agent to agent via shared
memory~\citep{gu2024agentsmith}; vertically, from individual users to
organizational memory; or temporally, from one session to another through
persistent profiles. It captures the risk that a poisoned memory may escape
its original context and become reused, inherited, or trusted across broader interaction scopes.
\ding{187} \textbf{Forget \& Rollback}. This phase concerns the system's ability to remove contaminated memories, restore an admissible historical state, trace the provenance of poisoned entries, and evaluate remediation. Snapshots and version histories support restoration, while forensic traceback helps localize contamination~\citep{zhang2025ragforensics}.
The memory lifecycle decomposition is also motivated in part by an analogy with human memory, as introduced in Appendix~\ref{appendix:humanmemory}.

\paragraph{Security objectives across phases.}
The second axis distinguishes four overlapping objectives. \textit{Integrity} concerns whether writes and transformations preserve the information and authority that a later decision may rely on. An unauthorized edit can corrupt a record, and an authorized summarizer can turn a tentative user claim into an apparent fact. \textit{Confidentiality} concerns inappropriate disclosure or inference across principals and channels. \textit{Availability} concerns access to legitimate memory and continuity after disruption or repair. \textit{Governance} concerns accountable control over admission, derivation, access, recovery, and deletion. Governance is a cross-cutting control objective rather than a fourth disjoint failure class. A graph-merge attack can undermine integrity at \textsc{Store} and \textsc{Execute}; aggressive filtering can reduce availability; restoration can recover useful state while conflicting with a subsequent erasure request. A work may therefore occupy several phase--objective pairs without requiring a new lifecycle stage. The following sections use phases as the narrative structure and objectives to identify overlapping failure types and control needs; Table~\ref{tab:new-evidence} maps representative claims to both axes and identifies the evidence used to test them.

\paragraph{What persists and how it is used.}
A fact and a skill can both be stored as text, although the skill may be compiled from several execution traces and later loaded as a procedure. How an auditor traces or corrects either kind of memory depends on its representation. A text summary or graph edge can be inspected alongside its sources. In MemT-1.0, however, each update modifies a recurrent matrix while the learned parameters remain fixed~\citep{memtensor2026memt1}; correcting a fact in such a representation requires evidence linking particular updates to its later recall.

KVUse illustrates another arrangement: a recommender updates a user profile and stores its key--value (KV) state for reuse across different request prefixes~\citep{guo2026kvuse}. In a reasoning-cache system, R$^2$ECAP finds that reusing cached state and recomputing it from the same visible text can yield different decisions because the cached state carries information from its construction context~\citep{zhu2026recap}. An audit of reused state should therefore identify its construction inputs and reader model. In inference-time experiments, KV quantization alters refusal behavior even when perplexity changes little~\citep{xu2026alignmentkv}. When a system retains compressed state across sessions, comparing behavior after reload against a full-precision control would test whether this effect persists.

\section{Attack Patterns Across the Lifecycle}\label{sec:attacks}

We organize the attack literature along the lifecycle introduced in
Section~\ref{sec:framework}, tracing how a poisoned memory is seeded at
\textsc{Write}, persists through \textsc{Store},
reactivates at \textsc{Retrieve}, steers behavior at
\textsc{Execute}, spreads via \textsc{Share \& Propagate}, and survives incomplete
cleanup during \textsc{Forget \& Rollback}.

\subsection{Write: Seeding Poisoned Memory}\label{ssec:attacks-write}

Within the memory lifecycle, the \textsc{Write} phase is where memory-poisoning attacks first enter the system.
AgentDojo evaluates indirect injection during tool use~\citep{debenedetti2024agentdojo}, and RAG poisoning benchmarks evaluate interference with retrieved evidence~\citep{zhang2025rsb}. A memory-enabled agent can retain the result of an encounter and use it again when the original input is gone.
Table~\ref{tab:write-threat} groups representative write-path attacks by attacker access, ranging from corpus writes to query-only interaction and influence over external observations. These settings expose different routes into persistent state.

Representative works illustrate these access settings.
AgentPoison~\cite{chen2024agentpoison}
demonstrates that a corpus-level adversary can plant embedding-space
backdoors with under 0.1\% poison rate and over 80\% attack success.
MINJA~\citep{dong2025minja} demonstrates that query-only interaction suffices to inject persistent malicious records into memory.
eTAMP~\citep{zou2026etamp} extends the threat to environment-only influence: an attacker can manipulate a web page the agent encounters during normal operation without interacting directly with the agent.
Together, these results point to a clear defensive implication: \textit{the effective write-stage attack surface includes not only explicit memory updates, but also any observable context that can influence what the agent decides to store}.

Hidden in Memory~\citep{pulipaka2026hiddenmemory} extends this pattern to sleeper instructions embedded in external documents: the agent stores apparently user-relevant content that can steer a later session. Its main protocol separates successful writing from subsequent use, so a conditional activation rate is not an end-to-end attack rate. MPBench~\citep{dash2026mpbench} examines four write channels and includes skill-procedure insertion, where a task trajectory becomes a reusable malicious procedure. These results connect exposure at \textsc{Write} to consolidation at \textsc{Store}, including persistent skills as well as factual records.

\subsection{Store: Provenance, Retention, and Audit in Long-Term Memory}
The \textsc{Store} phase is a critical control point in long-term memory security because it determines how written entries are indexed, compressed, retained, and evicted before later retrieval or execution. 
First, indexing determines which future queries can reactivate a memory
entry, making retrieval keys, semantic labels, and storage-priority assignments 
potential adversarial amplifiers~\cite{tian2026injecmem,chen2024agentpoison}.
Second, retention, decay, and eviction policies jointly regulate both availability and confidentiality. Retention schemes based on access frequency or recency may inadvertently keep adversarial entries alive while discarding legitimate ones. Furthermore, without automatic expiration or user-accessible deletion, sensitive content can accumulate in memory indefinitely, creating growing privacy exposure over time~\citep{zhong2024memorybank,li2025memos-full,zhang2024ghost,zou2026etamp}.
Third, provenance, write logs, snapshots, and diff-auditable histories support complementary forensic and recovery tasks~\citep{li2025memos-full,zhang2025ragforensics}. Their requirements depend on the implementation: whole-memory snapshot rollback, for example, need not rely on entry-level semantic lineage~\citep{su2026chronomem}. These mechanisms motivate storage-time governance alongside retrieval- and execution-time defenses.

MemForm~\citep{pan2026memform} further examines the interfaces through which persistent memories are formed: content, evidence, scope, selection, and consolidation path. Under its controlled memory protocols, downstream behavior can change when the retained experiences or their consolidation order change, even when individual experiences remain valid. Even a single claim can acquire greater certainty as it is written into memory. In PASB, a user supplies a claim in one session and the agent is later tested in a fresh session; stored notes sometimes express the claim more confidently than the user did, and editing those notes reduces the tested agents' subsequent reliance on it~\citep{mao2026pasb}.

ShadowMerge~\citep{luo2026shadowmerge} makes graph consolidation itself an attack surface. Ordinary interactions introduce conflicting relations that extraction, entity merging, and relation normalization place near the same graph anchor, increasing their chance of being retrieved together. GraphRAG Under Fire~\citep{liang2025graphrag} poisons a source corpus with text designed to alter extracted relations. ShadowMerge studies interaction-driven updates to persistent agent memory, emphasizing how newly extracted relations are merged with existing graph state in its evaluated Mem0 setting.

\subsection{Retrieve and Execute: From Selection to Action Steering}
\label{ssec:attacks-retrieve}

\textsc{Retrieve} is the chokepoint that reconnects past writes with present use, allowing a memory entry written weeks earlier to re-enter the active context window and influence the agent's current behavior.
MCFA~\citep{xu2026mcfa} shows that a sufficiently salient retrieved memory can override the user's explicit instruction and dictate the tools the agent invokes, their invocation order, and their arguments, highlighting the risk that poisoned memory can usurp the agent's operational autonomy.
MemoryGraft~\citep{srivastava2025memorygraft} extends this retrieval-to-execution threat beyond explicit tool control by showing that poisoned procedural memories, once retrieved as prior ``successful experiences,'' can be reused as task-solving strategies and silently replace the agent's behavior without a separate trigger. A reused procedure can also enter the evidence from which the agent learns its next skill. In a multi-round skill-distillation study, removing the harmful source skills recovered little of the lost performance, while an oracle that also removed all of their descendants recovered more, though some loss persisted~\citep{shang2026skillcontamination}. Propagation can therefore occur even within a single agent, as one procedure shapes which procedures emerge in later sessions.
MemMorph~\citep{zhang2026memmorph} targets tool selection through memories optimized to survive rewriting and match anticipated future queries. Its direct-backend and indirect-dialogue settings distinguish write privileges, while optimization uses knowledge of the memory writer and a shadow-agent template. This supplies a concrete \textsc{Store}--\textsc{Retrieve}--\textsc{Execute} chain in which rewritten experience biases which available tool the agent selects.

Taken together, the \textsc{Write}, \textsc{Store}, \textsc{Retrieve},
and \textsc{Execute} phases yield the end-to-end attack chain.
A manipulated web observation encountered during routine browsing can be summarized as a memory entry and persisted to long-term storage.
Days or weeks later, in a different session and possibly a different
task context, that entry is retrieved and silently steers tool
invocation.
In this delayed-use chain, the initial poisoning window closes before later execution begins, so single-turn detection cannot observe both ends of the chain.

Continuing reinforcement provides another persistence path. Zombie Agents studies how subsequent memory updates can renew an injected payload and sustain unauthorized behavior across sessions~\citep{yang2026zombie}.

\begin{table*}[t]
\centering
\small
\setlength{\tabcolsep}{4.2pt}
\renewcommand{\arraystretch}{1.15}
\caption{Representative write-path attacks grouped by attacker access. Success rates retain each work's protocol and denominator and are not directly comparable.}
\vspace{-0.5em}
\label{tab:write-threat}
\begin{tabularx}{\textwidth}{
@{}
>{\raggedright\arraybackslash}p{2.45cm}
>{\raggedright\arraybackslash}p{2.0cm}
>{\raggedright\arraybackslash}p{2.45cm}
>{\centering\arraybackslash}p{1.35cm}
>{\raggedright\arraybackslash}X
@{}}
\toprule
\rowcolor{green!8}
\textbf{Work}
& \textbf{Attacker Access}
& \textbf{Activation / Use Condition}
& \makecell{\textbf{Cross}\\\textbf{Session}}
& \textbf{Mechanism and Evidence} \\
\midrule

AgentPoison\newline\citep{chen2024agentpoison}
& Corpus write
& Optimized backdoor
& \checkmark
& \textit{Embedding-space poisoning.} Optimizes poisoned memories with ${<}0.1\%$ poison rate and ${\geq}80\%$ ASR. \\ \grayline

\addlinespace[2pt]

InjecMEM\newline\citep{tian2026injecmem}
& Single interaction
& Topic-conditioned
& \checkmark
& \textit{Retriever-agnostic anchoring.} Combines anchors with Multi-GCG: 76.6\% ASR-c conditional on retrieval and 35.6\% joint ASR on MemoryOS. \\ \grayline

\addlinespace[2pt]

MINJA\newline\citep{dong2025minja}
& Query-only
& Victim-term retrieval
& \checkmark
& \textit{Self-generated poison.} Uses bridging steps and indication prompts to make the agent create the malicious memory. \\ \grayline

\addlinespace[2pt]

MemoryGraft\newline\citep{srivastava2025memorygraft}
& Ingestion artifact
& Trigger-free
& \checkmark
& \textit{Procedural-memory grafting.} Imitates semantic procedures rather than injecting factual memories. \\ \grayline

\addlinespace[2pt]

eTAMP\newline\citep{zou2026etamp}
& Environment only
& Observation-based
& \checkmark
& \textit{Web-observation poisoning.} Adversarial web content is absorbed during browsing, producing $8\times$ ``frustration'' amplification. \\
\grayline
\addlinespace[2pt]
Hidden in Memory\newline\citep{pulipaka2026hiddenmemory}
& External document
& Later-session use
& \checkmark
& \textit{Sleeper-memory poisoning.} Separately evaluates memory writing and later behavioral activation. \\

\bottomrule
\end{tabularx}
\vspace{-1.5em}
\end{table*}

\subsection{Share \& Propagate: Cross-Session Influence and Disclosure}
\label{ssec:attacks-share}

In multi-agent and shared-state systems, contamination can spread through inter-agent messages, shared memory stores, and tool-mediated data flows, where poisoned content is passed as arguments to external tools or written into shared artifacts.
This section covers propagation, extraction, membership inference, and unintended disclosure during ordinary personalization.
\citet{gu2024agentsmith} demonstrate infectious jailbreak in simulated populations of up to one million LLaVA-1.5 agents using randomized pairwise chats.
\citet{men2024contagious} demonstrate
analogous propagation through dialogue;
and ComPromptMized~\citep{cohen2024morrisii} constructs
zero-click self-replicating prompts that traverse GenAI-powered email
assistants, with each recipient's agent absorbing and re-emitting the
payload to its own contacts.

\paragraph{Agent-generated propagation.}
\citet{das2026selfpropagating} study agents whose misaligned goals or values are induced through system prompts. They can encode those goals in persistent state for a later agent, including a different model, to pursue. When the dedicated memory interface is disabled, some evaluated agents pass those goals on through shared files.

\paragraph{Extraction, membership, and unintended disclosure.}
ADAM~\citep{lyu2026adam} estimates the distribution of private memory from observed responses to adapt its extraction queries. Its ASR is the fraction of attempts that recover at least one previously unseen record; database coverage and query budget remain separate quantities. MRMMIA~\citep{chen2026mrmmia} instead tests whether a known candidate belongs to chat-agent memory, aggregating multiple recall probes under output-only, token-probability, or retrieval-trace access. Membership disclosure can matter even when an attacker cannot reconstruct unknown records.

MemLeak~\citep{mudgal2026memleak} traces cross-user disclosure to pooled retrieval that returns another user's entries. In its synthetic-user experiments, partitioning the store, filtering by ownership metadata, or checking ownership on the returned context can each block this path. Even when an agent draws on the right user's history, its replies can elicit further disclosure. In multi-session simulations, \citet{kaneko2026disclosure} find that agents repeat or presuppose personal attributes, prompting replies that reinforce the stored profile. The feedback loop runs from \textsc{Execute}, where the response uses the profile, back to \textsc{Write}, where the user's reply can strengthen it.

\subsection{Forget \& Rollback: Residual State and Failed Recovery
}
The \textsc{Forget} phase removes poisoned or sensitive memory state following attack detection, correction, or an authorized deletion request.
Its central risk is incomplete removal: information remains available after the system has applied the requested intervention.
Incomplete forgetting arises because a single memory item may leave derivatives across raw dialogue logs, summarized memory cards, vector indexes, reflected lessons, shared stores, and audit records \citep{packer2023memgpt,xu2025amem,chhikara2025mem0,li2025memos-full,rezazadeh2025collaborative}. Deleting only the visible entry can therefore leave retrievable residues in compressed summaries, retrieval indices, or propagated copies, allowing contamination to reappear after apparent cleanup \citep{wang2024ragunlearning,wang2026sbu}.
MemCut~\citep{zhang2026memcut} distinguishes native recovery through ordinary retrieval from residual recovery when a reader is directly given surviving objects linked to annotated disclosures or evidence. These separate access conditions show why reduced retrieval success alone is insufficient evidence of forgetting; the benchmark tests controlled textual evidence trails rather than unrestricted recoverability.

Execution-state unlearning~\citep{yao2026executionstate} extends the residual-state question to active summaries, plans, and cached execution state after a persistent record is removed. AfterDeleteBench~\citep{saleh2026afterdeletebench} separately measures reconstruction, reinsertion into memory, and recall in a later session using exposed-then-deleted and never-exposed controls. Its synthetic-string and controlled-derivative tasks distinguish residual disclosure from the additional effect attributable to prior exposure.

The request and access conditions also matter. Bemula~\citep{xie2026bemula} evaluates natural-language forget requests with paired controls, measuring requested removal, preservation of unrelated memories, relearning, and extraction resistance. AUBench~\citep{lin2026aubench} frames unlearning across parametric knowledge, persistent memory, and tool-mediated acquisition. These are complementary protocols: deleting a stored record, reconstructing a clean execution state, and preventing reacquisition through an accessible tool require different interventions.

\section{Defenses: Prevention, Containment, and Recovery}\label{sec:defense}

Defenses intervene at different stages of the memory lifecycle, combining prevention, containment, and recovery.
At \textsc{Write} time, the central objective is to control how externally sourced content is admitted, labeled, and subsequently used. VerificAgent~\cite{nguyen2025verificagent} introduces human-verified freezing, where candidate memories are audited before being committed to a frozen safety contract. MemCube-style memory units~\citep{li2025memos-full} instead attach source, version, sensitivity, and temporal metadata at write time, making later audit, access control, and rollback possible. 
At \textsc{Store} time, versioned snapshots, write logs, content-addressable records, and compression audits provide architectural support for different recovery tasks. Provenance standards, such as W3C PROV~\citep{provdm2013}, further offer a principled serialization layer for representing and managing these lineage-bearing memory objects. MemLineage~\citep{ouyang2026memlineage} connects signed provenance and derivation records to a sensitive-action gate. Its propagation guarantee requires the critical derivation edges to remain above an attribution threshold.

TMA-NM~\citep{louck2026tmanm} studies how summarization, trusted-tool echoes, and manufactured corroboration can launder apparent authority. It binds authority to authenticated origins and explicit endorsement, relying on the implementation to label and attribute origins correctly. Louck checks this rule in a bounded TLA+ model and supplements the checks with an inductive-invariant argument. The rule links provenance to action policy across \textsc{Store} and \textsc{Execute}.

Among the memory lifecycle stages, the \textsc{Retrieve} phase has seen the most concentrated development of defensive mechanisms.
RobustRAG~\citep{xiang2024robustrag} uses isolate-then-aggregate generation to certify query-specific lower bounds on response quality under a bounded number of corrupted retrieval passages.
TrustRAG~\citep{zhou2025trustrag} and SeCon-RAG~\citep{si2025seconrag} use clustering, trust scoring, semantic similarity, and contradiction-aware filtering to distinguish benign from malicious retrieved content. 
RevPRAG~\citep{tan2025revprag} detects poisoning through activation differences between poisoned and non-poisoned generations. 
A-MemGuard~\citep{wei2025amemguard} is more memory-native: it retrieves semantically related memories, compares independent reasoning paths, treats disagreement as an anomaly signal, and distills detected attacks into lesson memories for future hardening. 
\citet{karunanidhi2026utilityattack} exposes a complementary limit through plain false assertions that evade the evaluated content screening. In the studied similarity regime, stronger provenance penalties can recover accuracy but can also exclude legitimate evidence from untrusted sources. This connects integrity protection to the availability of useful memory and the need for external grounding.

Between retrieval and action, three decisions require different evidence. First, an entry must be eligible for the present request: a former policy can be relevant to a historical audit yet ineligible as guidance for today's task. \citet{wang2026admissibility} compare retrieval routes at matched recall of required evidence and track which entries reach the prompt. Second, entries that pass individually can conflict when assembled. MemGov checks a bounded candidate set against supplied eligibility and pairwise constraints, returning a compatible set or a record of the conflicting constraints~\citep{christou2026memgov}. Third, the action layer must establish that the requested operation is permitted and its prerequisites hold. For decisions that may skip an already completed check, Proof-of-Execution Memory consults a receipt written by the trusted execution layer~\citep{rahman2026poem}.

At \textsc{Execute}, CaMeL, FIDES, and PCAS enforce information-flow control by separating data and control planes and restricting the influence of untrusted content on tool use or program flow~\citep{debenedetti2025camel,costa2025fides,palumbo2026pcas}. Progent, IsolateGPT, and type-directed privilege separation further reduce blast radius through fine-grained tool policies, execution isolation, and typed privilege boundaries~\citep{shi2025progent,wu2024isolategpt,jacob2025typeprivilege}.

\textsc{Share}-time defenses shift the focus from content-level inspection to principal-aware access and interaction modeling.
Collaborative Memory~\citep{rezazadeh2025collaborative} models multi-user and multi-agent access through a time-evolving principal--resource graph, while BlindGuard~\citep{miao2025blindguard} uses unsupervised graph-level anomaly detection to flag previously unseen multi-agent contagion patterns. 
MemLeak~\citep{mudgal2026memleak} evaluates explicit ownership enforcement on retrieved context, complementing isolation earlier in the retrieval pipeline. PerMIT~\citep{jiang2026permit} mediates environment-derived values through a finite memory interface and exact-value, tool-scoped confirmations. The Lean proof covers action integrity in the single-user core and reasoning integrity for a fixed display transcript. The extension to community endorsements relies on additional probabilistic assumptions.

When a private entry is promoted to shared memory, a second agent may acquire access to evidence collected for a different task. MAPLE-Guard checks this transition as well as the private write, retrieval, and subsequent cross-agent reuse~\citep{xiong2026mapleguard}. Zep Agent Skills traces the fields of a versioned skill to the runs that support them and withdraws dependent skills from search pending repair if a cited trace is deleted~\citep{chalef2026skills}. Publishing a skill to another namespace in the same security domain submits it to destination-side admission: its traces explain why the procedure was learned, whereas the receiving system must decide whether the new principal may run it for the task.

\paragraph{Privacy across use and disclosure.}
MemPrivacy~\citep{chen2026memprivacy} places sensitive-span substitution and the restoration map on the local device, enabling cloud-side processing with typed placeholders. EP-Mem~\citep{sun2026epmem} instead carries user-configured disclosure policies across memory storage, retrieval, and generation, including audience-specific exceptions. These mechanisms address data minimization and contextual release in addition to ownership. Response-side suppression of personal-attribute statements also reduces the feedback loop observed by \citet{kaneko2026disclosure} in their simulations while retaining much of the measured response-quality benefit. Retrieval authorization and appropriate downstream disclosure are therefore related but distinct checks.

\paragraph{Recovery targets and conditions.}
Finally, mechanisms and evaluations for the \textsc{Forget \& Rollback} phase address distinct recovery objectives.
RAGForensics~\citep{zhang2025ragforensics} shows that post-hoc traceback is feasible for static RAG corpora, and \citet{owaspmemoryguard2026} provide an industry reference for integrity checks, policy enforcement, snapshots, and rollback primitives. 
ChronoMem~\citep{su2026chronomem} supports semantic selection and restoration of whole-memory snapshots, evaluating rollback-consistent behavior after exposure to later updates; its scope excludes arbitrary external side effects. MemSecBench~\citep{chen2026memsecbench} evaluates persistence, downstream consequences, and selective repair through a \textsc{Write}--\textsc{Execute}--\textsc{Forget} protocol, with \textsc{Execute} and \textsc{Forget} branching independently from the same verified post-\textsc{Write} state. These complementary results motivate evaluating how recovery composes with provenance and access control across subsequent memory updates.

Dependency-guided rollback repair~\citep{wang2026rollbackrepair} traces memory-to-action dependencies, preserves independently supported state, and selectively recomputes affected steps after a fault is diagnosed. Its improvements in immediate recovery and cost coexist with residual recurrence. Execution-state unlearning~\citep{yao2026executionstate} restores a clean prefix and replays a sanitized suffix under deterministic or fixed-randomness, replayable-runtime assumptions. A counterfactual repair preserving later valid work has a different target from restoring a historical snapshot; both require separate handling of already committed external effects.

In LinEra, the reach of a deletion depends on which sources were visible to the writer during derivation, all of which are recorded as dependencies. Under its cryptographic assumptions, deleting a source makes the protected contents of its recorded descendants unreadable even to an adversary who later obtains the store and current keys~\citep{ameur2026linera}. On ten LoCoMo conversations with a Qwen2.5-7B writer, forgetting 1\% of turns erased 73.0\% of derived memories when the writer summarized whole sessions, compared with 3.8\% when it used four-turn windows. On downstream question answering before re-derivation, F1 fell from 27.8 to 13.8 for session summaries versus 31.8 to 30.7 for four-turn windows, reflecting what the surviving memories could still support.

\section{Verifiable Memory Governance: An Operational Framework}\label{sec:sovereignty}

The lifecycle analysis identifies five capabilities for maintaining auditable, recoverable control over memory. We organize them as \textbf{Verifiable Memory Governance (VMG)}:
\begin{itemize}[leftmargin=*]

\item \textbf{Write Authorization (WA)} requires that every entry committed to long-term memory be attributable to an authenticated source and pass an explicit authorization check before consolidation, preventing untrusted external content from entering memory as if user-endorsed.

\item \textbf{Provenance Visibility (PV)} requires that every memory entry carry a queryable, lineage-complete provenance record tracing back to its originating write event through any intermediate summarization or merging steps, enabling post-breach attribution and forensic analysis.

\item \textbf{Principal-Scoped Retrieval (PS)} requires that the retrieval function return only entries whose current authorized scope includes the querying principal, restricting cross-principal disclosure through retrieval.
\item \textbf{Rollbackability (RB)} requires retained recovery points sufficient to restore memory under current erasure and access-control constraints, supporting recovery after a successful attack.
\item \textbf{Verified Forgetting (VF)} requires a specified bound on post-deletion exposure of target content within declared memory substrates, access channels, and probing distributions.
\end{itemize}

As summarized in Table~\ref{tab:sovereignty} in Appendix~\ref{app:primitives}, for each primitive, we specify a representative threat, a corresponding control direction, and proposed evaluation metrics.

\paragraph{Formal specification.}
We express the requirements as predicates over memory state $M_t$ at time $t$. An evaluator first specifies the system boundary, interfaces, policies, lineage granularity, and test oracles.
Each entry \(m \in M_t\) carries metadata\,%
\((\operatorname{src}(m),\allowbreak
  \operatorname{scope}(m),\allowbreak
  \operatorname{prov}(m),\allowbreak
  \operatorname{ver}(m))\)
denoting source principal, current authorized principal set, provenance
record, and version identifier.
$M_t$ denotes the memory state within that boundary, including the logs, summaries, indices, snapshots, and controlled copies covered by the claim. This includes controlled files, active working summaries, pending plans, and cached execution state when they carry the target information~\citep{das2026selfpropagating,yao2026executionstate}.

\paragraph{Audit and retrieval interfaces.}
$\mathrm{Queryable}(m)$ means that an authorized audit principal can retrieve and resolve $\operatorname{prov}(m)$ for version $\operatorname{ver}(m)$. $\mathrm{LineageComplete}(m)$ means that this record identifies the originating write events and all parent-version dependencies required by the specified transformation model. Completeness is checked through the authorized audit view against instrumented or controlled traces. Lower-privilege provenance views remain access-controlled and may be redacted. Dependency records need not retain deleted source payloads, and any retained audit data remain subject to the stated erasure policy.

$R$ denotes an entry-returning retrieval interface before synthesis or assembly. Derived summaries or graph aggregates require a separate release check over their contributing entries and applicable disclosure policy.

\paragraph{Recovery view.}
Let $L(M)$ denote the live-memory view to be restored, including the logical indices and caches needed to serve it. Equality is evaluated at this logical-view granularity, not the physical layout of caches or indices. Append-only audit history and recovery-point management are not rewound by the RB equality. Retained logs, snapshots, and controlled copies remain subject to the erasure and access requirements within the system boundary.
Formally, we define the following criteria:
\vspace{-0.5em}
\begin{align}
\textbf{WA}(M_t)
  &:=
  \begin{aligned}[t]
  &\forall m \in M_t,\;
    \operatorname{src}(m) \neq \bot \\
  &{}\land
    \operatorname{auth_W}\!\left(
      \operatorname{src}(m),m,t_w(m)
    \right)=1,
  \end{aligned}
  \label{eq:wa}
\intertext{where $\bot$ denotes a missing source principal. A successful $\operatorname{auth_W}$ check authenticates the source, binds it to the committed write event, and permits that write under the policy at $t_w(m)$. It is an evaluated policy predicate, not a self-reported approval flag.}
\textbf{PV}(M_t)
  &:=
  \begin{aligned}[t]
  &\forall m \in M_t,\;
    \mathrm{Queryable}(m) \\
  &{}\land\,
    \mathrm{LineageComplete}(m),
  \end{aligned}
  \label{eq:pv}
\intertext{with queryability and completeness evaluated through the audit interface and transformation model defined above.
}
\textbf{PS}(M_t)
  &:=
  \begin{aligned}[t]
  &\forall q,\pi,\;
    R(q,\pi,M_t) \\
  &{}\subseteq
    \left\{
      m \in M_t :
      \pi \in \operatorname{scope}(m)
    \right\},
  \end{aligned}
  \label{eq:ps}
  \intertext{which requires that, for any query $q$ and requesting principal $\pi$,
the output of $R(q,\pi,M_t)$ be strictly restricted to entries for
which $\pi \in \operatorname{scope}(m)$.} 
\textbf{RB}(M_t)
  &:=
  \begin{aligned}[t]
  &\forall\, t_0 \in \mathcal{T}_t, \\
  &L\!\left(\operatorname{rollback}(M_t,t_0)\right) \\
  &\qquad =P_t\!\left(L(M_{t_0})\right),
  \end{aligned}
  \label{eq:rb}
  \intertext{where $\mathcal{T}_t$ is a declared, nonempty set of retained and permitted recovery points before $t$. $P_t$ applies current erasure and access-control constraints to the historical live-memory view, preserving subsequent deletion and revocation decisions.
  }
\textbf{VF}_{\varepsilon}(M_t)
  &\notag\\
  &:=
  \begin{aligned}[t]
  &\forall X \in \mathcal{X}, \\
  &\Pr_{q \sim Q_X}
    \!\left[
      \operatorname{Expose}_X
      \!\left(q,F_X(M_t)\right)=1
    \right] \\
  &{}\le \varepsilon,
  \end{aligned}
  \label{eq:vf}
  \end{align}
which bounds exposure for deletion targets $\mathcal{X}$. $F_X$ is the deletion operation, $Q_X$ specifies the probing queries or query sequences, and $\operatorname{Expose}_X$ tests reconstruction through the accessible channels. A finite probe suite estimates this probability for the stated scope.
An evaluation must use a nonempty target family, a policy-selected $0\leq\varepsilon<1$ fixed before testing, and declared access conditions and query budgets. Pre-deletion or matched no-deletion controls establish that the targets were retained or accessible before the intervention.

A system satisfies $\varepsilon$-VMG at time $t$, under these policies and evaluation conditions, if all five hold simultaneously:
\begin{equation}\label{eq:ms}
\textbf{VMG}_{\varepsilon}(M_t)
  := \textbf{WA}\!\land\!\textbf{PV}\!\land\!\textbf{PS}
     \!\land\!\textbf{RB}\!\land\!\textbf{VF}_{\varepsilon}.
\end{equation}
\begin{figure}[t]\centering
\input{contents/vmg-primitives-figure}
\caption{Five governance capabilities acting on a shared memory state. Connections identify their roles in memory operations.}
\label{fig:sovereignty-tower}
\end{figure}

\paragraph{Useful service under these constraints.}
Deployment evaluation pairs the governance constraints with a nontrivial, memory-dependent authorized workload $\mathcal{D}$, including required benign writes and later reads whose answers are not supplied by model priors or unrestricted tool reacquisition. The utility measure and minimum acceptable score $u_{\min}$ are specified before testing. Let $U_{\mathcal{D}}$ denote the evaluated agent's performance on $\mathcal{D}$, and $U_{\mathcal{D}}^{\emptyset}$ its performance with all memory channels relevant to that workload disabled and the same permitted tools. Acceptance requires $U_{\mathcal{D}}\geq u_{\min}>U_{\mathcal{D}}^{\emptyset}$. Legitimate-write acceptance and authorized-read success are reported separately. The workload respects current erasure decisions and requires retention of other admissible information.

Figure~\ref{fig:sovereignty-tower} places these capabilities around their shared object: the evolving memory state. Authenticated write records support attribution; provenance helps locate derivatives for repair or deletion; principal scoping controls retrieval. Rollback and forgetting coordinate restoration with removal. Their implementation relationships depend on the architecture, with erasure and revocation constraints linking recovery decisions across time. Together, the predicates provide an operational specification for system design and evaluation.

\paragraph{From memory admission to authorized action.}
WA checks a write against the admission policy, whose adequacy depends on the threat model. Using the stored value in an action also calls for checking its factual basis and the permission to act. PV preserves the derivation record needed to apply action policy across transformations. Origin-bound authority in TMA-NM and exact-value confirmations in PerMIT provide concrete mechanisms at this boundary~\citep{louck2026tmanm,jiang2026permit}.

\paragraph{Evaluating a change in project membership.}
Consider a project assistant that records a restricted release configuration for a collaborator and uses it to draft handover notes. The write must be authorized (WA), and an auditor must be able to trace the note to the source versions and any access decision it cites (PV). Drawing on CoSec's changes in project membership and GateMem's tests of requester-specific access~\citep{chen2026cosec,ren2026gatemem}, we propose comparing three continuations of the same history: the collaborator keeps access, loses it, or loses and then regains it. To locate the first divergence between histories, we track what retrieval returns to the collaborator (PS), what reaches the model, and what the assistant discloses or does.

Once access is revoked, the next handover should exclude the configuration, yet a superseded policy can still guide the assistant's action~\citep{shen2026revoked}. Reinstatement tests the converse failure, in which the assistant continues to deny an authorized release because it relies on an earlier denial~\citep{zhanglin2026decisions}. A separate branch tests what happens when the configuration is deleted and an older snapshot is restored. Under RB, the restored live view should match the historical live view with current erasures and access rules applied; under VF, running the same probes over accessible derivatives before deletion, after erasure, and after restoration estimates exposure and shows at which checkpoints the configuration can be recovered. Finally, a release requiring a different, still-authorized configuration tests continuity of service.

\filbreak
\section{Conclusion and Future Work}

As LLM agents increasingly rely on persistent memory~\cite{packer2023memgpt,chhikara2025mem0,li2025memos-full}, cross-session long-term memory introduces risks beyond conventional input-centric or single-turn security frameworks. We propose a Memory Lifecycle Framework organizing attacks and defenses across six phases and four objectives, showing that memory threats unfold as cross-phase chains where poisoned content can persist, propagate, and resist cleanup. We further introduce Verifiable Memory Governance, specifying five primitives for auditable and recoverable memory control, and highlight key gaps in provenance infrastructure, cross-principal propagation, and post-deletion verification.

How should an agent use its history after a fact, a permission, or the representation of a memory changes? The project example in Section~\ref{sec:sovereignty} suggests holding the prior history and legitimate task fixed, varying one change, and tracing its effect on the evidence loaded and the work completed. Clean-history baselines cover recall across conversations in LongMemEval and LoCoMo~\citep{wu2024longmemeval,maharana2024locomo} and operations on evolving histories in MemBench, MemoryAgentBench, and MemOps~\citep{tan2025membench,hu2026memoryagentbench,hao2026memops}. Personalized tasks also require deciding when a preference applies~\citep{yoon2026benchpres} and checking whether personal context encourages an unsafe response~\citep{guo2026psbench}. Table~\ref{tab:new-evidence} gives the comparison conditions for security claims, including how the interaction history was constructed. For example, MemoryDojo tracks retrieval of both the original item and its derivatives, and in its cross-domain evaluation adding benign interactions changed how often poisoning reached later tasks~\citep{yang2026memorydojo}.

To quantify the consequences of a changed decision, counterfactual memory regret compares downstream task loss under matched clean and edited memory states~\citep{zou2026memharm}. HEAMS evaluates a single configuration across separate tasks for utility, disclosure, user control, and security; in its disclosure evaluation, a recall policy reduced leakage while also lowering coverage of information the agent was permitted to share~\citep{goel2026heams}, illustrating why both permitted and prohibited outcomes must be tracked. We propose following repaired systems through subsequent updates to measure whether faults recur, whether authorized work resumes, and how repair costs compare with re-reading current permitted sources.

\section*{Limitations}
This review synthesizes publicly documented systems through database search, citation expansion, and targeted updates. Its coverage emphasizes architectures and mechanisms rather than an exhaustive census of the literature.

VMG specifies requirements over declared memory states, interfaces, and policies. Deployment-level assurance requires evaluating how these controls compose across memory, tools, and principals. Forgetting results remain tied to the evaluated access channels and probe distributions.

\section*{Acknowledgments}
This work was supported by the Shanghai Special Fund for Promoting High-Quality Industrial Development: Pilot Industry Innovation and Development Program (Artificial Intelligence Track, ID: 2025-GZL-RGZN-02006).

\begingroup
\emergencystretch=1em
\bibliography{references}
\endgroup

\clearpage
\appendix
\emergencystretch=1em

\twocolumn[{
    \renewcommand\twocolumn[1][]{#1}
    \begin{center}
      \textbf{\fontsize{15}{48}\selectfont Appendix}
    \end{center}
    \vspace{0.5cm}
}]

\section{Architectural Governance Primitives}
\label{app:primitives}
Table~\ref{tab:sovereignty} summarizes the five VMG primitives from Section~\ref{sec:sovereignty}, specifying for each the representative threat, corresponding defense direction, and proposed evaluation metric.

\begin{table*}[t]
\centering
\caption{Architectural governance primitives.}
\label{tab:sovereignty}
\footnotesize
\setlength{\tabcolsep}{4pt}
\begin{tabularx}{\textwidth}{@{}>{\raggedright\arraybackslash}p{2.0cm}LLL@{}}
\toprule
\rowcolor{green!8}
\textbf{Primitive} &
\textbf{Representative Threat} & \textbf{Control Direction} &
\textbf{Proposed Metric} \\
\midrule
Write Authorization (WA) &
Query-induced injection; environment-injected poisoning~\citep{dong2025minja,tian2026injecmem,zou2026etamp} &
Typed write validation; human or policy arbitration~\citep{nguyen2025verificagent,jacob2025typeprivilege,palumbo2026pcas} &
Adaptive injection-survival rate at bounded false-positive rate (FPR)~\citep{nasr2025adaptive} \\
\addlinespace

Provenance Visibility (PV) &
Source-monitoring failure; imitation-based grafting~\citep{srivastava2025memorygraft} &
Provenance-gated retrieval; lineage-preserving compression~\citep{provdm2013,souza2025provagent} &
Fraction of entries traceable to source event \\
\addlinespace

Principal-Scoped Retrieval (PS) &
Cross-user contamination; black-box extraction~\citep{yang2026ucc,wang2025mextra,elyagoubi2026agentleak,he2025edea} &
Capability- or label-based retrieval policy~\citep{li2025aac,costa2025fides,mireshghallah2025cimemories}; ownership enforcement~\citep{mudgal2026memleak} &
Cross-principal leakage rate under scripted query suites \\
\addlinespace

Rollbackability (RB) &
Compression-amplified toxins; behavioral drift~\citep{packer2023memgpt,xu2025amem} &
Versioned restoration; forensic localization~\citep{su2026chronomem,zhang2025ragforensics} &
Restoration accuracy under current policies; recovery coverage and latency \\
\addlinespace

Verified Forgetting (VF) &
Residual derivatives; reappearance in summaries~\citep{zhang2024ghost,pulipaka2026persistbench} &
Deletion coordinated over declared substrates and derivatives~\citep{bourtoule2021machine,maini2024tofu,wang2024ragunlearning,wang2026sbu} &
Post-deletion exposure under declared access and probe conditions~\citep{maini2024tofu,wang2024ragunlearning,zhang2026memcut,saleh2026afterdeletebench} \\
\bottomrule
\end{tabularx}
\end{table*}

Use-time endorsement and origin-bound action authority complement admission and provenance controls~\citep{jiang2026permit,louck2026tmanm}. Selective dependency repair complements historical restoration but has a different recovery target~\citep{wang2026rollbackrepair}.

\section{Human Memory as a Conceptual Analogy}
\label{appendix:humanmemory}
Human memory is reconstructive: retrieval can alter a memory, and a person can remember a claim while misremembering its source~\citep{squire2015memory,schacter1999cognitive,schacter2007constructive,nader2000fear,zaki2025engram}. Studies of misinformation and social contagion document how a report from another person can enter later recall~\citep{loftus2005planting,roediger2001social}. These observations motivate a question for agent memory: which distinctions survive when an experience is recalled and written again?

Agent architectures implement retention and reconstruction through specific state updates~\citep{zhang2026metis}. An externally supplied claim can reappear as the agent's own experience~\citep{dong2025minja,tian2026injecmem}, as a lesson distilled from several episodes~\citep{packer2023memgpt,xu2025amem}, or as another agent's input~\citep{gu2024agentsmith,men2024contagious}. For these cases, source attribution, transformation history, and destination scope identify different places to inspect the chain.

\section{Methodology for Paper Selection}

The original review searched ACL Anthology, the ACM Digital Library, IEEE Xplore, USENIX proceedings, OpenReview, and arXiv for work on long-term, episodic, and agent memory, poisoning, injection, extraction, propagation, governance, forgetting, and rollback. Reference and citation tracing through Google Scholar, arXiv, and Semantic Scholar expanded the pool. The original review screened more than 200 candidates and retained approximately 90. For this initial search, we recorded the databases and keyword families, whereas exact query strings, search cutoffs, and item-level screening decisions were not archived.

Targeted updates on 23 and 30 October and through 19 November 2026 selected distinct mechanisms, representations, policy boundaries, and evaluation methods. We traced references and citation alerts, compared candidates with the existing bibliography, and checked each addition against its primary text. We cite the relevant proceedings record or preprint version.

\section{VMG Operationalization}
For Equations~\eqref{eq:wa}--\eqref{eq:ms}, declare the system boundary and policy before applying these checks.
\begin{itemize}[leftmargin=*]
    \item WA: compare authenticated write events and their recorded decisions against the declared policy. Use published injection attacks (e.g., MINJA, eTAMP) to measure injection survival and false positives.
    \item PV: under designated audit credentials, resolve provenance for each entry version and compare it with controlled or instrumented dependency traces. Test lower-privilege queries for prohibited metadata disclosure. Include summarization, merging, and tool echoes; separately test whether transformations improperly increase action authority~\citep{louck2026tmanm}.
    \item PS: issue entry-returning retrieval queries under different principal identities and measure results exceeding the requester's authorized scope. Test release of synthesized or aggregated outputs separately under the applicable disclosure policy.
    \item RB: declare the live-memory projection and recovery points, trigger rollback, and compare the restored view with $P_t(L(M_{t_0}))$, including current erasure and revocation constraints. Check reconstruction of required indices and caches, and preservation of permitted audit history. Report restoration accuracy, coverage, and latency separately.
    \item VF: use tiered probes (direct, paraphrased, associative, multi-turn) to test reconstruction under declared access paths and judge criteria, reporting targets, accessible derivatives, probe counts, and uncertainty. Report reconstruction, reinsertion, later-session recall, preservation, and relearning separately~\citep{saleh2026afterdeletebench,xie2026bemula}; declare whether tool reacquisition is permitted and distinguish it from surviving memory~\citep{lin2026aubench}. K-Bench extends observation across agent channels in separate weight, prompt, and retrieval settings~\citep{yu2026kbenchunlearning}.

\end{itemize}

\paragraph{Paired deletion audits.}
AfterDeleteBench pairs an exposed-then-deleted state with a matched never-exposed control~\citep{saleh2026afterdeletebench}. Their exposure difference estimates the incremental effect of prior exposure; it does not replace the absolute bound in VF, since both states can have high disclosure.

\section{Real-System Analysis}
We compare the interfaces described in Mem0 (2504.19413v1) and MemOS (2507.03724v1)~\citep{chhikara2025mem0,li2025memos-full}.
\begin{itemize}[leftmargin=*]
    \item Mem0 describes LLM-selected ADD, UPDATE, DELETE, and NOOP operations (Section 2.1). Its graph variant, Mem0$^g$, records creation timestamps and retains invalidated relations for temporal reasoning (Section 2.2). These support tracking memory changes; the cited version leaves dedicated write-authorization and rollback protocols unspecified.
\item MemOS describes provenance and version metadata, read/write/share scopes, snapshots, historical rollback, and lifecycle policies (Sections 4.2, 5.3.2, 5.4.3, and 5.5). These provide implementation hooks for VMG. Evaluating them requires testing lineage completeness across transformations, authorization under adversarial updates, and erasure of derivatives and propagated copies.

\end{itemize}

Bemula complements these mechanism descriptions with measurements tied to the tested implementation, model, and request protocol~\citep{xie2026bemula}.

\paragraph{Author relationships.}
MemPrivacy and MAPLE-Guard share authors Zhiyu Li and Feiyu Xiong with this survey. MemOS, MemOps, Metis, MemT-1.0, and the memory-value paper also share authors with it. Zehao Lin is a coauthor of the working paper on decision reuse and authorization recovery.

\section{Revision Notes}
\label{app:revision}
The October 2026 update clarified the four objectives and the live-memory projection in Equation~\eqref{eq:rb}. The November 2026 update adds the authorization diagnostic and the evidence comparison in Table~\ref{tab:new-evidence}.

\clearpage
\onecolumn
\section{Evidence for Lifecycle Security Claims}
\label{app:evidence-map}
\begin{minipage}{\linewidth}
\centering
\small
\setlength{\tabcolsep}{4pt}
\renewcommand{\arraystretch}{1.08}
\captionsetup{type=table,justification=raggedright,singlelinecheck=false}
\caption{Evidence for representative lifecycle claims. The second column gives the phases and main objectives: integrity (I), confidentiality (C), availability (A), and governance (G).}\label{tab:new-evidence}
\begin{tabular}{@{}>{\raggedright\arraybackslash}p{.205\textwidth}>{\raggedright\arraybackslash}p{.155\textwidth}>{\raggedright\arraybackslash}p{.273\textwidth}>{\raggedright\arraybackslash}p{\dimexpr.367\textwidth-6\tabcolsep\relax}@{}}
\toprule
\textbf{Study} & \textbf{Phases and objectives} & \textbf{What is observed} & \textbf{Comparison or condition} \\
\midrule
\multicolumn{4}{@{}l}{\textbf{Empirical evaluations}}\\*
\addlinespace
MemoryDojo \citep{yang2026memorydojo} & Write, Store, Retrieve, Execute (I) & Which item is retrieved, which later tasks are affected, and whether any task in a trajectory is affected. & Snapshots distinguish original from subsequently contaminated items. The history-prefixed setting also measures competition from benign memory.\\
\addlinespace
MemHarm \citep{zou2026memharm} & Write, Execute (I) & Downstream task loss attributable to an edited memory. & Pair clean and edited states on the same task and loss function. Severity comparisons depend on the admissible edit class.\\
\addlinespace
CoSec \citep{chen2026cosec} and GateMem \citep{ren2026gatemem} & Retrieve, Execute, Share, Forget (C, A, G) & CoSec checks actions and artifacts after a community changes. GateMem checks permitted use, access violations, and responses after deletion. & CoSec scores against the policy in force at action time. GateMem uses requester-specific checkpoints in a shared history and observes the agent-facing interface.\\
\addlinespace
MRMMIA \citep{chen2026mrmmia} & Retrieve, Execute (C) & Membership inference for a known candidate memory. & Specify whether the observer sees only outputs, token probabilities, or retrieval traces. The inference target is membership of the candidate.\\
\addlinespace
MemSecBench \citep{chen2026memsecbench} & Write, Execute, Forget (I, A) & Persistence, subsequent action, and selective repair with useful memory retained. & Execution and repair branch separately from the same verified post-write state.\\
\addlinespace
MemCut \citep{zhang2026memcut} and AfterDeleteBench \citep{saleh2026afterdeletebench} & Forget, Retrieve, Write (C, G) & Recovery through ordinary retrieval or surviving evidence; reconstruction, reinsertion, and later recall. & MemCut varies reader access. AfterDeleteBench compares exposed-then-deleted and never-exposed histories. Record absolute exposure alongside the difference.\\
\addlinespace
ChronoMem \citep{su2026chronomem} & Store, Rollback (A, G) & Restoration of a selected historical snapshot and behavior consistent with that state. & Compare with the chosen historical target after intervening updates. Undoing external actions requires a separate mechanism.\\
\addlinespace
HEAMS \citep{goel2026heams} & Write, Retrieve, Forget (C, A, G) & Task utility, allowed information coverage, inappropriate disclosure, and compliance with edit or deletion requests. & A single configuration is evaluated across separate offline tasks; deletion probes measure response behavior.\\
\midrule
\multicolumn{4}{@{}l}{\textbf{System guarantees and their scope}}\\*
\addlinespace
MemGov \citep{christou2026memgov} & Retrieve (I, G) & A compatible working set or an auditable constraint conflict before assembly. & The certificate covers the bounded candidates and supplied constraints. Missing evidence or undetected contradictions remain upstream.\\
\addlinespace
PerMIT \citep{jiang2026permit} & Store, Execute (I, G) & Action integrity and reasoning integrity under the specified interface. & Lean proofs cover the single-user core, with reasoning integrity conditioned on the displayed transcript. Community trust has separate probabilistic assumptions.\\
\addlinespace
LinEra \citep{ameur2026linera} & Store, Forget (C, A, G) & Readability of protected descendant contents, fraction of derived objects erased, and retained QA utility. & The storage guarantee assumes recorded dependencies and compromise of current keys after erasure. The writer's input window affects the cost of erasing descendants.\\
\bottomrule
\end{tabular}
\end{minipage}

\end{document}